\documentclass[aps,pra,twocolumn]{revtex4-1}
\usepackage{tabularx}
\usepackage{dcolumn}
\usepackage{graphics}% Include figure files
\usepackage{bm}% bold matha
\usepackage[dvips]{graphicx}
\usepackage[dvips]{rotating}
\usepackage[latin1]{inputenc}
\usepackage{dcolumn}
\usepackage{tabularx}
\usepackage{amsmath}
\usepackage{amssymb}
\usepackage{float}
\usepackage{natbib}
\usepackage[dvipsnames]{xcolor}
\begin{document}
\draft

\title{Ultracold collisions of Cs in excited hyperfine and Zeeman states}

\author{Matthew D. Frye}
\affiliation{Joint Quantum Centre (JQC) Durham-Newcastle, Department of Chemistry, Durham University, South Road, Durham, DH1 3LE, United Kingdom.}
\author{B. C. Yang}
\affiliation{Joint Quantum Centre (JQC) Durham-Newcastle, Department of Chemistry, Durham University, South Road, Durham, DH1 3LE, United Kingdom.}
\author{Jeremy M. Hutson}
\email{j.m.hutson@durham.ac.uk}
\affiliation{Joint Quantum Centre (JQC) Durham-Newcastle, Department of Chemistry, Durham University, South Road, Durham, DH1 3LE, United Kingdom.}

\date{\today}

\begin{abstract}
We investigate Cs+Cs scattering in excited Zeeman and hyperfine states. We
calculate the real and imaginary parts of the s-wave scattering length; the
imaginary part directly provides the rate coefficient for 2-body inelastic
loss, while the real part allows us to identify regions of magnetic field where
3-body recombination will be slow. We identify field regions where Cs in its
$(f,m_f)=(3,+2)$ and $(3,+1)$ states may be stable enough to allow
Bose-Einstein condensation, and additional regions for these and the $(3,0)$
and $(3,-3)$ states where high-density clouds should be long-lived.
\end{abstract}

\maketitle

\section{Introduction}

The ability to cool atoms to ultracold temperatures has opened up a huge field
of physics over the past 3 decades. A key feature of ultracold atoms is the
ability to control the interatomic interactions by varying the scattering
length $a$. This is most commonly done using a zero-energy Feshbach resonance,
where a bound state crosses a threshold as a function of magnetic field $B$
\cite{Chin:RMP:2010}. In the absence of inelastic scattering, there is a
resonant pole in $a(B)$ \cite{Moerdijk:1995} which allows essentially any
scattering length to be obtained with sufficiently good field control.

The scattering length shows very different behavior for different atomic
species. For the alkali metals, which are particularly commonly used, every
stable isotope exhibits Feshbach resonances at accessible magnetic fields.
However, widths and background scattering lengths vary enormously, making each
isotope suitable for a different range of experiments \cite{Chin:RMP:2010}.
Indeed, even different Zeeman and hyperfine states of the same isotope have
different properties and may find different applications.

A pair of alkali-metal atoms in their $^2$S electronic ground state may
interact on singlet ($^1\Sigma_g^+$) or triplet ($^3\Sigma_u^+$) potential
curves. Each of these is characterized by a single (field-independent)
scattering length, $a_\textrm{s}$ and $a_\textrm{t}$ respectively. Different
Zeeman and hyperfine states experience different combinations of the singlet
and triplet interactions and have Feshbach resonances at different fields. In
general terms, Feshbach resonances due to s-wave states crossing threshold are
narrow if $a_\textrm{s} \approx a_\textrm{t}$ but may be broad otherwise. The
alkali-metal atoms with the broadest resonances and therefore the most
precisely tunable scattering lengths are $^6$Li \cite{Houbiers:1998,
Jochim:2002, Julienne:Li67:2014}, $^{39}$K \cite{dErrico:2007}, and $^{133}$Cs
\cite{Vuletic:1999, Chin:2000, Leo:2000, Chin:cs2-fesh:2004,
Berninger:Cs2:2013}.

Cs has a very large positive triplet scattering length $a_\textrm{t} =
2858(19)\,a_0$ and a moderate positive singlet scattering length $a_\textrm{s}
= 286.5(10)\,a_0$ \cite{Berninger:Cs2:2013}. For its lowest Zeeman state,
($f,m_f)=(3,+3)$, there are many resonances, some of which are very broad. The
broad resonances provide excellent control over the scattering length, making
Cs an attractive atom for studies of strongly interacting Bose gases
\cite{Chevy:2016} and Efimov physics \cite{Kraemer:2006, Berninger:Efimov:2011,
Huang:2nd-Efimov:2014}. Mixtures of Cs with other species are also of interest,
particularly for studying systems with large mass imbalances
\cite{Fratini:2012, Tung:2014, Gopalakrishnan:2015, Ardila:2016, DeSalvo:2019}
and for heteronuclear molecule formation \cite{Brue:AlkYb:2013, Patel:2014,
Molony:RbCs:2014, Takekoshi:RbCs:2014}. However, the intraspecies scattering
length for Cs (3,+3) is very large at fields away from resonance, causing fast
3-body recombination \cite{Esry:1999} at most magnetic fields and making it
challenging to work with high Cs densities. In particular, it has been possible
to cool Cs (3,+3) close to degeneracy only at a few specific magnetic fields;
it is usually done at the Efimov minimum in 3-body recombination near 21~G
\cite{Kraemer:2006}, but it is also possible around at 558.7 and 894~G
\cite{Berninger:Cs2:2013}.

The dependence of the scattering length on magnetic field is well known for Cs
(3,+3). Feshbach resonance positions and near-threshold bound-state energies at
fields up to 1000~G have been fitted to obtain precise singlet and triplet
potential curves, and the calculated scattering length has been tabulated for
magnetic fields up to 1200~G \cite{Berninger:Cs2:2013}. In addition, a
considerable amount of early work used Cs in its magnetically trappable states
$(3,-3)$ and (4,+4) \cite{Arndt:1997, Guery-Odelin:1998, Soding:1998,
Vuletic:1999, Hopkins:2000, Thomas:2003}, while Chin \emph{et al.}\
\cite{Chin:2000, Chin:cs2-fesh:2004} observed Feshbach resonances in a variety
of states and in mixtures at magnetic fields up to 230~G. These were
interpreted to obtain interaction potentials \cite{Leo:2000,
Chin:cs2-fesh:2004}. However, relatively little has been done on the excited
states since Bose-Einstein condensation was achieved in the (3,+3) state
\cite{Weber:CsBEC:2003}, and the interaction potentials of Refs.\
\cite{Leo:2000} and \cite{Chin:cs2-fesh:2004} do not predict resonance
positions accurately at higher fields \cite{Berninger:Cs2:2013}. There is a
clear need for a thorough investigation of the collisional properties of Cs in
excited Zeeman and hyperfine states, using the most recent interaction
potential \cite{Berninger:Cs2:2013}. Excited Cs atoms may provide new species
with new Feshbach resonances and additional regions of stability. This may be
particularly valuable for mixture experiments where interspecies resonances
appear at specific fields \cite{Brue:AlkYb:2013, Patel:2014, Yang:CsYb:2019},
or where the second atom itself imposes limitations on the fields that can be
used \cite{Cho:RbCs:2013, Patel:2014}.

\section{Theory}

We perform coupled-channel scattering calculations on the interaction potential
of \citeauthor{Berninger:Cs2:2013} \cite{Berninger:Cs2:2013}. The methods used
are similar to those in Ref.\ \cite{Berninger:Cs2:2013}, so only a brief
outline is given here. The Hamiltonian for the interacting pair is
\begin{equation}
\label{full_H}
\hat{H} =\frac{\hbar^2}{2\mu}\left[-\frac{1}{R}\frac{d^2}{dR^2}R
+\frac{\hat{L}^2}{R^2}\right]+\hat{H}_\textrm{A}+\hat{H}_\textrm{B}+\hat{V}(R),
\end{equation}
where $R$ is the internuclear distance, $\mu$ is the reduced mass, and $\hbar$
is the reduced Planck constant. $\hat{L}$ is the two-atom rotational angular
momentum operator. The single-atom Hamiltonians $\hat{H}_i$ contain the
hyperfine couplings and the Zeeman interaction with the magnetic field.
The interaction operator $\hat{V}(R)$ contains the two isotropic
Born-Oppenheimer potentials, for the X $^1\Sigma_g^+$ singlet and $a$
$^3\Sigma_u^+$ triplet states, and anisotropic spin-dependent couplings which
arise from dipole-dipole and second-order spin-orbit coupling.

Scattering calculations are carried out using the \textsc{molscat} package
\cite{molscat:2019}. The scattering wavefunction is expanded in a fully
uncoupled basis set that contains all allowed spin functions, limited by
$L_{\rm max}=4$. The collision energy is $E=1\ \textrm{nK}\times
k_\textrm{B}$. Solutions are
propagated from $R_\textrm{min}=6\, a_0$ to $R_\textrm{mid}=20\, a_0$ using the
diabatic modified log-derivative propagator of Manolopoulos
\cite{Manolopoulos:1986} with a step size of 0.002 $a_0$, and from
$R_\textrm{mid}$ to $R_\textrm{max}=10000\, a_0$ using the log-derivative Airy
propagator of Alexander and Manolopoulous \cite{Alexander:1987} with a variable
step size. The log-derivative matrix is transformed into the
asymptotic basis set at $R_\textrm{max}$ and matched to S-matrix boundary
conditions to obtain the scattering matrix {\bf S}.

For collisions of Cs in its lowest state (3,+3), only elastic collisions are
possible. For collisions of atoms in excited states, however, inelastic
scattering may occur. Inelastic collisions that produce atoms in lower-lying
states release kinetic energy and usually produce heating or trap loss. The
inelastic collisions are of two types: spin exchange and spin relaxation.
Collisions that conserve $M_F=m_{f,\textrm{A}} + m_{f,\textrm{B}}$ are termed
spin-exchange collisions, while those that do not conserve $M_F$ are termed
spin-relaxation collisions. Spin-exchange collisions are driven mostly by the
difference between the singlet and triplet interactions, whereas
spin-relaxation collisions are driven by the much weaker anisotropic couplings.

Spin-exchange collisions are generally fast when they are
energetically allowed. However, for pairs of alkali-metal atoms in the lower
hyperfine state ($f=3$ for Cs), they are endoergic. The incoming and outgoing
channels have the same linear Zeeman energy, but are separated by terms that
are quadratic in $B$ at low field. At 10~G, spin-exchange collisions are
allowed for collision energies above 130~$\textrm{nK} \times k_{\rm B}$.
Below this threshold, only spin relaxation can produce inelasticity.

Both elastic and inelastic collisions are conveniently characterized in terms
of the energy-dependent s-wave scattering length \cite{Hutson:res:2007},
\begin{equation}
a(k) = \frac{1}{ik} \left(\frac{1-S_{00}}{1+S_{00}}\right),
\end{equation}
where $k=\sqrt{2\mu E}/\hbar$ is the wavevector and $S_{00}$ is the diagonal
S-matrix element in the incoming channel. When there is only one open channel,
only elastic scattering is possible and $a(k)$ is real. When inelastic
scattering is possible, however, $a(k)$ is complex, $a(k) = \alpha(k) - i
\beta(k)$. The
rate coefficient for 2-body loss is \cite{Hutson:res:2007}
\begin{equation}
k_2=\frac{4\pi\hbar\beta}{\mu(1+k^2|a|^2+2k\beta)}.
\label{eq:k2}
\end{equation}
This expression for $k_2$ can contain small contributions from s-wave
collisions that change $L$ without changing the internal state of the atoms,
but these vanish as $E \rightarrow 0$ and are negligible at the collision
energies considered here. When the denominator of Eq.\ \ref{eq:k2} can be
neglected, $\beta=1\ a_0$ corresponds to $k_2=1.3 \times 10^{-12}$
cm$^3$\,s$^{-1}$. Both the scattering length and the loss rate are independent
of energy in the limit $E\rightarrow0$. Deviations from this reach around 2\%
at $E=100\ \textrm{nK} \times k_\textrm{B}$, but are negligible at the energy
of our calculations. Inelastic rates from s-wave collisions generally decrease
with energy except as discussed below for spin-exchange collisions. The height
of the d-wave centrifugal barrier is $180\ \mu\textrm{K} \times k_\textrm{B}$,
and d-wave contributions to $k_2$ are generally small at collision energies
below $50\ \mu\textrm{K} \times k_\textrm{B}$, except near narrow resonances.

In the presence of inelastic scattering, the real and imaginary parts of the
scattering length show an oscillation rather than a pole. The amplitude of the
oscillation is characterized by the resonant scattering length $a_\textrm{res}$
\cite{Hutson:res:2007}. If the background inelastic scattering is negligible,
$\alpha(B)$ shows an oscillation of amplitude $\pm a_\textrm{res}$ and $\beta$
shows a peak of magnitude $a_\textrm{res}$. For the states of Cs considered
here, $\alpha$ is typically several hundred $a_0$ or more, while $\beta$ is
often in the range $0 < \beta < 10\ a_0$. It is therefore common for decayed
resonances to be visible in plots of $k_2$ (obtained from $\beta$) but not in
the corresponding plots of $\alpha$.

\section{Results}

\begin{figure*}[htbp]
\includegraphics[width=0.95\textwidth]{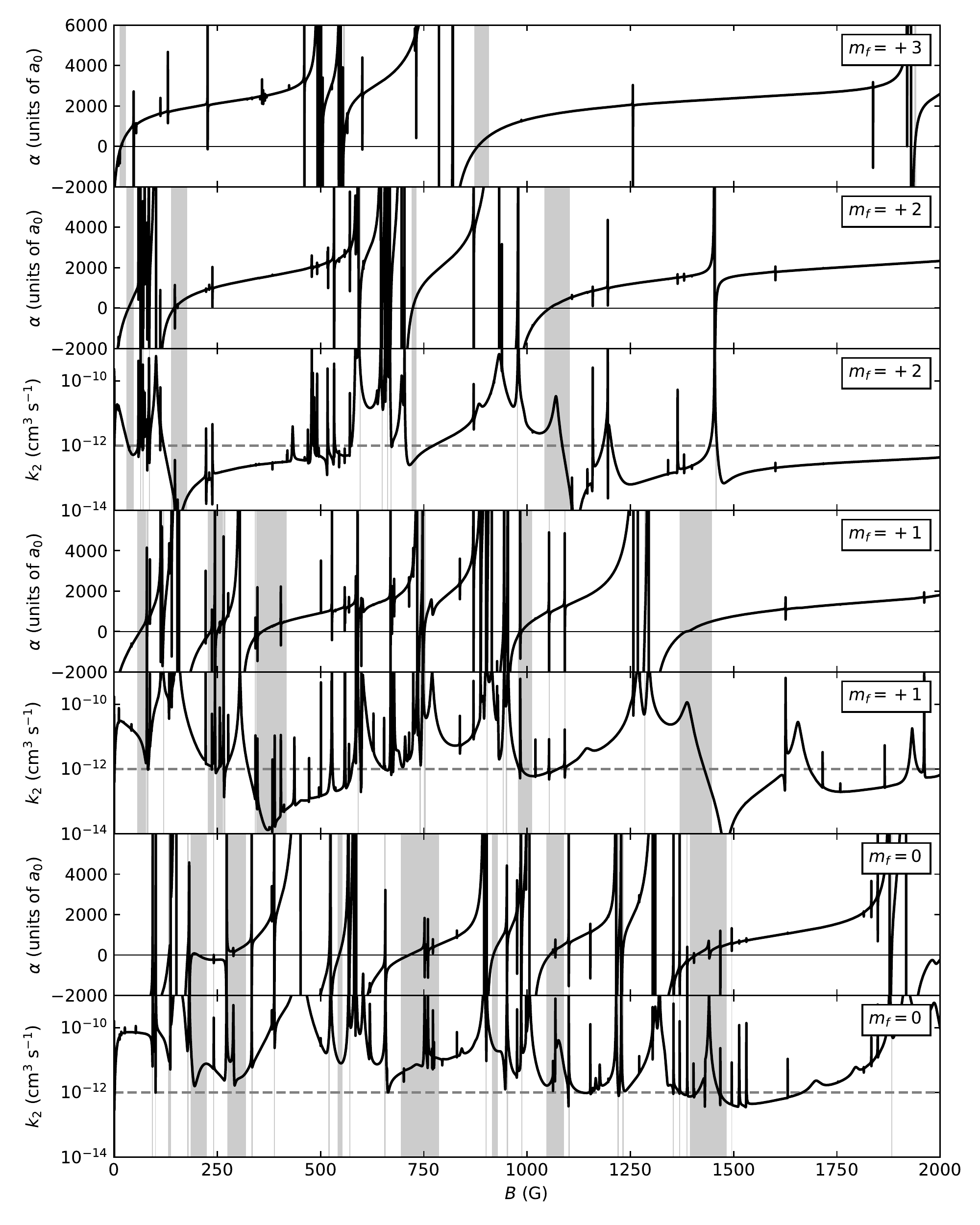}
\caption{Real part of the scattering length $\alpha$ and inelastic loss rate
coefficient $k_2$ for collisions of pairs of Cs atoms with $f=3$ and the same
$m_f$, for $m_f\geq0$. Shaded regions correspond to fields where
$-200\,a_0<\alpha<500\,a_0$. Calculations are performed on a 0.1~G grid, so
narrow resonances are not always visible. \label{fig:Cs2_3_p}}
\end{figure*}

\begin{figure*}[htbp]
\includegraphics[width=0.95\textwidth,clip=true,trim=0 0 0 3.3cm]{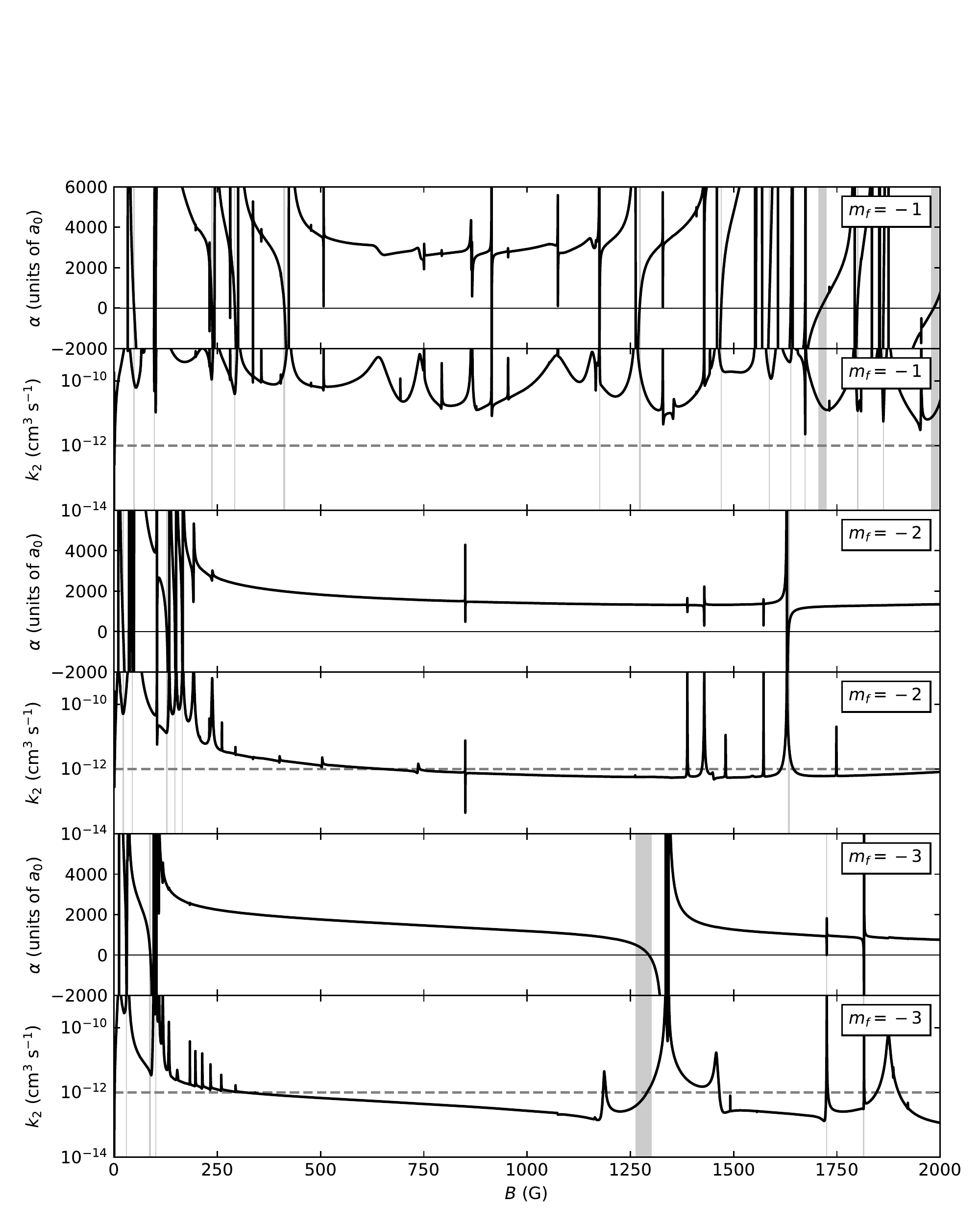}
\caption{Real part of the scattering length $\alpha$ and inelastic loss rate
coefficient $k_2$ for collisions of pairs of Cs atoms with $f=3$ and the same
$m_f$, for $m_f < 0$. Shaded regions correspond to fields where
$-200\,a_0<\alpha<500\,a_0$. Calculations are performed on a 0.1~G grid, so
narrow resonances are not always visible. \label{fig:Cs2_3_m}}
\end{figure*}

3-body recombination rates depend strongly on the scattering length
\cite{Esry:1999}. For Cs, 3-body losses are generally fast except in limited
ranges of magnetic field near broad resonances, where either the scattering
length is near a zero-crossing or 3-body losses are suppressed by an Efimov
minimum \cite{Kraemer:2006}. Evaporative cooling is most efficient near an
Efimov minimum, since it requires elastic collisions and the elastic cross
section vanishes at a zero-crossing. For broad resonances in Cs, Efimov minima
typically occur when $a\sim200$ to $300\,a_0$ \cite{Berninger:Cs2:2013}.

Before Bose-Einstein condensation was achieved for Cs (3,+3)
\cite{Weber:CsBEC:2003}, degeneracy was approached but not achieved for Cs
$(3,-3)$. Cooling of $(3,-3)$ was limited by 2-body inelastic collisions with a
rate coefficient around $2 \times 10^{-12}$ cm$^3$\,s$^{-1}$ at 139~G
\cite{Thomas:2003}. We therefore estimate that a rate coefficient higher than
about $10^{-12}$ cm$^3$\,s$^{-1}$ is sufficient to prevent cooling to
degeneracy in other states. It may be noted that the scattering length at 139~G
is calculated below to be 3200~$a_0$; this is sufficient to cause substantial
3-body losses, which limited the Cs density in Ref.\ \cite{Thomas:2003}.

We have carried out coupled-channel scattering calculations for pairs of Cs
atoms initially in the same state ($f$,$m_f$) for all $f=3$ and $f=4$ states.
The calculations are carried out at fields from 0 to 2000~G in steps of 0.1~G.
We calculate the real and imaginary parts of the scattering length, $\alpha(B)$
and $\beta(B)$, and express the latter as the 2-body inelastic rate coefficient
$k_2(B)$.

Figures \ref{fig:Cs2_3_p} and \ref{fig:Cs2_3_m} show the results for $f=3$,
$m_f\geq0$ and $m_f<0$, respectively. The grey bars show where $-200\,a_0 <
\alpha < 500\,a_0$, to indicate regions where the 3-body recombination rate is
expected to be moderate. The bars serve to guide the eye in reading the
corresponding values of $k_2$.

The scattering length for $m_f=+3$ is known from previous work
\cite{Berninger:Cs2:2013} and will not be not discussed in detail here. The
regions of moderate scattering length around 21~G and 894~G, where cooling is
usually performed, are clearly visible; the region near 558.7~G is too narrow
to be clearly seen with our scale/grid.

For $m_f=+2$, the behavior of $\alpha$ is broadly similar to $m_f=+3$. There
are a few very broad resonances and a large number of narrower resonances.
Inelastic loss is now possible, and every resonance also creates a
corresponding peak in $k_2$. Many of these peaks are asymmetric and have a dip
in loss on one side that arises from interference between background inelastic
scattering and inelastic scattering mediated by the resonance
\cite{Hutson:res:2007, Hutson:HeO2:2009}. If there is a single dominant loss
channel, the interference may be almost complete, and $k_2$ then drops close to
zero. However, additional loss channels result in incomplete cancelation and
shallower minima. For the broad resonances, these dips can be quite wide and
tend to coincide with the regions of moderate scattering length. This results
in several ranges where both $\alpha$ and $k_2$ are small enough to allow
experiments with high densities of Cs. The region around 150~G associated with
the broad resonance at 100~G appears particularly promising. At 150~G,
spin-exchange collisions are allowed at collision energies above
25~$\mu\textrm{K} \times k_{\rm B}$. However, such
collisions actually \emph{reduce} the kinetic energy; the (3,+3) and (3,+1)
atoms produced will remain confined in an optical trap, and can return to the
original state in further collisions. The region below 1100~G, associated with
broad resonance near 930~G, is affected by a narrower resonance that is
strongly decayed, with $a_\textrm{res}=25\, a_0$, and is thus visible only in
the inelastic rate. Because of this, Cs (3,+2) is likely to exhibit slow 2-body
loss only at the upper end of this second shaded region, and the low vales of
$k_2$ must be balanced against 3-body recombination arising from the increasing
value of $\alpha$.

Similar effects are seen for $m_f=+1$. The shaded region between 350 and 400~G
is generally favorable, though it contains a number of narrow resonances that
will produce loss. The shaded region from 1370 to 1450~G is affected by a
narrow resonance that enhances 2-body loss at the lower end of the range, so
that Cs (3,+1) will probably exhibit slow 2-body loss only at the upper end of
this range.

The loss rates for $m_f=0$ and $-1$ show weaker resonant structure with
shallower troughs. Many of the resonances also appear as oscillations in
$\alpha$ rather than poles. This is due to the increased number of loss
channels. There are few regions with low 2-body loss rates, and these do not
coincide with moderate $\alpha$ except for a small region near 1500~G for
$m_f=0$.

There are significantly fewer resonances visible for $m_f=-2$ and $-3$ than for
the lower states. This is largely because there are fewer closed channels close
in energy to support resonant states. $\alpha$ therefore remains large across
the whole range for $m_f=-2$. The only broad resonance is at
1340~G for $m_f=-3$ and results in a region of moderate scattering length
around 1300~G. The inelastic rate coefficient in this region is about
$10^{-12}$~cm$^3$\,s$^{-1}$, so reasonably high-density clouds of Cs $(3,-3)$
might be stable.

\begin{figure*}[htbp]
\includegraphics[width=0.95\textwidth]{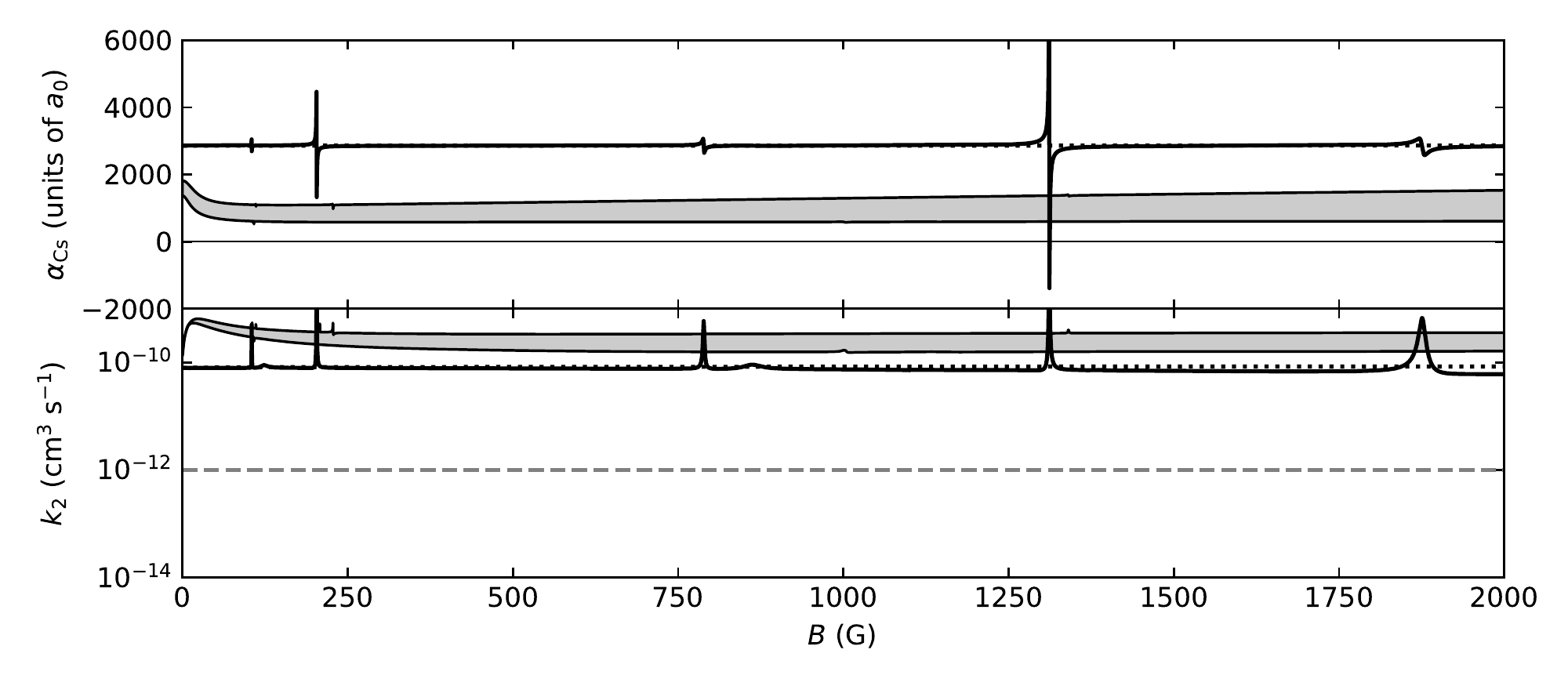}
\caption{Real part of the scattering length $\alpha$ and inelastic loss rate
coefficient $k_2$ for collisions of pairs of Cs atoms with $f=4$ and the same
$m_f$. The solid black line shows $m_f=-4$, while the dotted black line shows
$m_f=+4$ which is almost identical except near resonances. The shaded bands
show the ranges covered by $-3\leq m_f \leq +3$. Calculations are performed on
a 0.1~G grid, so narrow resonances are either not resolved. \label{fig:Cs2_4}}
\end{figure*}

Figure \ref{fig:Cs2_4} shows the results for all $f=4$ states. Both $k_2$ and
$\alpha$ are large over the entire range for all states. The lowest Zeeman
state, $m_f=-4$, exhibits the most structure, but there are only a few
resonances, which are narrow and significantly decayed. The remaining states
have little variation or structure because there are few closed channels at
higher energies to support resonant states, and the few resonances which do
exist are strongly decayed and so barely visible on this scale.

\section{Conclusions} \label{conclusion}

We have carried out coupled-channel calculations on collisions of
ultracold Cs in excited Zeeman and hyperfine states, in order to identify
regions of magnetic field where high-density atomic clouds might be cooled to
degeneracy or close to it. We have calculated the real and imaginary parts of
the scattering length at magnetic fields up to 2000~G for pairs of atoms in
each Zeeman and hyperfine state. The imaginary part of the scattering length
gives the rate coefficient for 2-body inelastic loss, while the real part
allows us to identify regions in which 3-body recombination will be relatively
slow.

For Cs in its $(f,m_f)$ = (3,+2) and (3,+1) excited states, there are regions
where the 2-body loss coefficient is very low, $k_2 \lesssim 10^{-14}$
cm$^3$\,s$^{-1}$, and 3-body loss is likely to be suppressed by Efimov effects.
These regions are very promising for creating high-density clouds and possibly
forming Bose-Einstein condensates. Cs (3,0) is less favorable, as the 2-body
loss coefficient seldom drops below $10^{-12}$~cm$^3$\,s$^{-1}$, but may
nevertheless offer possibilities. Cs $(3,-1)$ has even faster 2-body losses. Cs
$(3,-2)$ has large regions where the 2-body loss coefficient is slightly below
$10^{-12}$ cm$^3$\,s$^{-1}$, but the scattering length is large and there are
no broad Feshbach resonances in these regions to moderate 3-body losses.
$(3,-3)$ has a similar 2-body loss coefficient, but in this case there is a
broad Feshbach resonance that may produce low 3-body losses in a limited region
around 1300~G. All the Cs $f=4$ states experience fast 2-body losses across the
entire range of fields.

The calculations presented here open the way to producing high-density clouds
of Cs in excited Zeeman states with $f=3$. These are effectively new species
for the study of ultracold gases, with particular importance in studying atomic
mixtures and in heteronuclear molecule formation.

\begin{acknowledgments}
We are grateful to Simon Cornish and Alex Guttridge for valuable discussions.
This work was supported by the U.K. Engineering and Physical Sciences Research
Council (EPSRC) Grants No.\ EP/N007085/1, EP/P008275/1 and EP/P01058X/1.
\end{acknowledgments}

\bibliography{../all,Cs2_excited}

\begin{thebibliography}{38}%
\makeatletter
\providecommand \@ifxundefined [1]{%
 \@ifx{#1\undefined}
}%
\providecommand \@ifnum [1]{%
 \ifnum #1\expandafter \@firstoftwo
 \else \expandafter \@secondoftwo
 \fi
}%
\providecommand \@ifx [1]{%
 \ifx #1\expandafter \@firstoftwo
 \else \expandafter \@secondoftwo
 \fi
}%
\providecommand \natexlab [1]{#1}%
\providecommand \enquote  [1]{``#1''}%
\providecommand \bibnamefont  [1]{#1}%
\providecommand \bibfnamefont [1]{#1}%
\providecommand \citenamefont [1]{#1}%
\providecommand \href@noop [0]{\@secondoftwo}%
\providecommand \href [0]{\begingroup \@sanitize@url \@href}%
\providecommand \@href[1]{\@@startlink{#1}\@@href}%
\providecommand \@@href[1]{\endgroup#1\@@endlink}%
\providecommand \@sanitize@url [0]{\catcode `\\12\catcode `\$12\catcode
  `\&12\catcode `\#12\catcode `\^12\catcode `\_12\catcode `\%12\relax}%
\providecommand \@@startlink[1]{}%
\providecommand \@@endlink[0]{}%
\providecommand \url  [0]{\begingroup\@sanitize@url \@url }%
\providecommand \@url [1]{\endgroup\@href {#1}{\urlprefix }}%
\providecommand \urlprefix  [0]{URL }%
\providecommand \Eprint [0]{\href }%
\providecommand \doibase [0]{http://dx.doi.org/}%
\providecommand \selectlanguage [0]{\@gobble}%
\providecommand \bibinfo  [0]{\@secondoftwo}%
\providecommand \bibfield  [0]{\@secondoftwo}%
\providecommand \translation [1]{[#1]}%
\providecommand \BibitemOpen [0]{}%
\providecommand \bibitemStop [0]{}%
\providecommand \bibitemNoStop [0]{.\EOS\space}%
\providecommand \EOS [0]{\spacefactor3000\relax}%
\providecommand \BibitemShut  [1]{\csname bibitem#1\endcsname}%
\let\auto@bib@innerbib\@empty
%</preamble>
\bibitem [{\citenamefont {Chin}\ \emph {et~al.}(2010)\citenamefont {Chin},
  \citenamefont {Grimm}, \citenamefont {Tiesinga},\ and\ \citenamefont
  {Julienne}}]{Chin:RMP:2010}%
  \BibitemOpen
  \bibfield  {author} {\bibinfo {author} {\bibfnamefont {C.}~\bibnamefont
  {Chin}}, \bibinfo {author} {\bibfnamefont {R.}~\bibnamefont {Grimm}},
  \bibinfo {author} {\bibfnamefont {E.}~\bibnamefont {Tiesinga}}, \ and\
  \bibinfo {author} {\bibfnamefont {P.~S.}\ \bibnamefont {Julienne}},\
  }\href@noop {} {\bibfield  {journal} {\bibinfo  {journal} {Rev. Mod. Phys.}\
  }\textbf {\bibinfo {volume} {82}},\ \bibinfo {pages} {1225} (\bibinfo {year}
  {2010})}\BibitemShut {NoStop}%
\bibitem [{\citenamefont {Moerdijk}\ \emph {et~al.}(1995)\citenamefont
  {Moerdijk}, \citenamefont {Verhaar},\ and\ \citenamefont
  {Axelsson}}]{Moerdijk:1995}%
  \BibitemOpen
  \bibfield  {author} {\bibinfo {author} {\bibfnamefont {A.~J.}\ \bibnamefont
  {Moerdijk}}, \bibinfo {author} {\bibfnamefont {B.~J.}\ \bibnamefont
  {Verhaar}}, \ and\ \bibinfo {author} {\bibfnamefont {A.}~\bibnamefont
  {Axelsson}},\ }\href@noop {} {\bibfield  {journal} {\bibinfo  {journal}
  {Phys. Rev. A}\ }\textbf {\bibinfo {volume} {51}},\ \bibinfo {pages} {4852}
  (\bibinfo {year} {1995})}\BibitemShut {NoStop}%
\bibitem [{\citenamefont {Houbiers}\ \emph {et~al.}(1998)\citenamefont
  {Houbiers}, \citenamefont {Stoof}, \citenamefont {McAlexander},\ and\
  \citenamefont {Hulet}}]{Houbiers:1998}%
  \BibitemOpen
  \bibfield  {author} {\bibinfo {author} {\bibfnamefont {M.}~\bibnamefont
  {Houbiers}}, \bibinfo {author} {\bibfnamefont {H.~T.~C.}\ \bibnamefont
  {Stoof}}, \bibinfo {author} {\bibfnamefont {W.~I.}\ \bibnamefont
  {McAlexander}}, \ and\ \bibinfo {author} {\bibfnamefont {R.~G.}\ \bibnamefont
  {Hulet}},\ }\href {\doibase 10.1103/PhysRevA.57.R1497} {\bibfield  {journal}
  {\bibinfo  {journal} {Phys. Rev. A}\ }\textbf {\bibinfo {volume} {57}},\
  \bibinfo {pages} {R1497} (\bibinfo {year} {1998})}\BibitemShut {NoStop}%
\bibitem [{\citenamefont {Jochim}\ \emph {et~al.}(2002)\citenamefont {Jochim},
  \citenamefont {Bartenstein}, \citenamefont {Hendl}, \citenamefont
  {Hecker~Denschlag}, \citenamefont {Grimm}, \citenamefont {Mosk},\ and\
  \citenamefont {Weidem\"uller}}]{Jochim:2002}%
  \BibitemOpen
  \bibfield  {author} {\bibinfo {author} {\bibfnamefont {S.}~\bibnamefont
  {Jochim}}, \bibinfo {author} {\bibfnamefont {M.}~\bibnamefont {Bartenstein}},
  \bibinfo {author} {\bibfnamefont {G.}~\bibnamefont {Hendl}}, \bibinfo
  {author} {\bibfnamefont {J.}~\bibnamefont {Hecker~Denschlag}}, \bibinfo
  {author} {\bibfnamefont {R.}~\bibnamefont {Grimm}}, \bibinfo {author}
  {\bibfnamefont {A.}~\bibnamefont {Mosk}}, \ and\ \bibinfo {author}
  {\bibfnamefont {M.}~\bibnamefont {Weidem\"uller}},\ }\href {\doibase
  10.1103/PhysRevLett.89.273202} {\bibfield  {journal} {\bibinfo  {journal}
  {Phys. Rev. Lett.}\ }\textbf {\bibinfo {volume} {89}},\ \bibinfo {pages}
  {273202} (\bibinfo {year} {2002})}\BibitemShut {NoStop}%
\bibitem [{\citenamefont {Julienne}\ and\ \citenamefont
  {Hutson}(2014)}]{Julienne:Li67:2014}%
  \BibitemOpen
  \bibfield  {author} {\bibinfo {author} {\bibfnamefont {P.~S.}\ \bibnamefont
  {Julienne}}\ and\ \bibinfo {author} {\bibfnamefont {J.~M.}\ \bibnamefont
  {Hutson}},\ }\href@noop {} {\bibfield  {journal} {\bibinfo  {journal} {Phys.
  Rev. A}\ }\textbf {\bibinfo {volume} {89}},\ \bibinfo {pages} {052715}
  (\bibinfo {year} {2014})}\BibitemShut {NoStop}%
\bibitem [{\citenamefont {D'Errico}\ \emph {et~al.}(2007)\citenamefont
  {D'Errico}, \citenamefont {Zaccanti}, \citenamefont {Fattori}, \citenamefont
  {Roati}, \citenamefont {Inguscio}, \citenamefont {Modugno},\ and\
  \citenamefont {Simoni}}]{dErrico:2007}%
  \BibitemOpen
  \bibfield  {author} {\bibinfo {author} {\bibfnamefont {C.}~\bibnamefont
  {D'Errico}}, \bibinfo {author} {\bibfnamefont {M.}~\bibnamefont {Zaccanti}},
  \bibinfo {author} {\bibfnamefont {M.}~\bibnamefont {Fattori}}, \bibinfo
  {author} {\bibfnamefont {G.}~\bibnamefont {Roati}}, \bibinfo {author}
  {\bibfnamefont {M.}~\bibnamefont {Inguscio}}, \bibinfo {author}
  {\bibfnamefont {G.}~\bibnamefont {Modugno}}, \ and\ \bibinfo {author}
  {\bibfnamefont {A.}~\bibnamefont {Simoni}},\ }\href@noop {} {\bibfield
  {journal} {\bibinfo  {journal} {New Journal of Physics}\ }\textbf {\bibinfo
  {volume} {9}},\ \bibinfo {pages} {223} (\bibinfo {year} {2007})}\BibitemShut
  {NoStop}%
\bibitem [{\citenamefont {Vuleti\'c}\ \emph {et~al.}(1999)\citenamefont
  {Vuleti\'c}, \citenamefont {Kerman}, \citenamefont {Chin},\ and\
  \citenamefont {Chu}}]{Vuletic:1999}%
  \BibitemOpen
  \bibfield  {author} {\bibinfo {author} {\bibfnamefont {V.}~\bibnamefont
  {Vuleti\'c}}, \bibinfo {author} {\bibfnamefont {A.~J.}\ \bibnamefont
  {Kerman}}, \bibinfo {author} {\bibfnamefont {C.}~\bibnamefont {Chin}}, \ and\
  \bibinfo {author} {\bibfnamefont {S.}~\bibnamefont {Chu}},\ }\href@noop {}
  {\bibfield  {journal} {\bibinfo  {journal} {Phys. Rev. Lett.}\ }\textbf
  {\bibinfo {volume} {82}},\ \bibinfo {pages} {1406} (\bibinfo {year}
  {1999})}\BibitemShut {NoStop}%
\bibitem [{\citenamefont {Chin}\ \emph {et~al.}(2000)\citenamefont {Chin},
  \citenamefont {Vuleti\'c}, \citenamefont {Kerman},\ and\ \citenamefont
  {Chu}}]{Chin:2000}%
  \BibitemOpen
  \bibfield  {author} {\bibinfo {author} {\bibfnamefont {C.}~\bibnamefont
  {Chin}}, \bibinfo {author} {\bibfnamefont {V.}~\bibnamefont {Vuleti\'c}},
  \bibinfo {author} {\bibfnamefont {A.~J.}\ \bibnamefont {Kerman}}, \ and\
  \bibinfo {author} {\bibfnamefont {S.}~\bibnamefont {Chu}},\ }\href@noop {}
  {\bibfield  {journal} {\bibinfo  {journal} {Phys. Rev. Lett.}\ }\textbf
  {\bibinfo {volume} {85}},\ \bibinfo {pages} {2717} (\bibinfo {year}
  {2000})}\BibitemShut {NoStop}%
\bibitem [{\citenamefont {Leo}\ \emph {et~al.}(2000)\citenamefont {Leo},
  \citenamefont {Williams},\ and\ \citenamefont {Julienne}}]{Leo:2000}%
  \BibitemOpen
  \bibfield  {author} {\bibinfo {author} {\bibfnamefont {P.~J.}\ \bibnamefont
  {Leo}}, \bibinfo {author} {\bibfnamefont {C.~J.}\ \bibnamefont {Williams}}, \
  and\ \bibinfo {author} {\bibfnamefont {P.~S.}\ \bibnamefont {Julienne}},\
  }\href@noop {} {\bibfield  {journal} {\bibinfo  {journal} {Phys. Rev. Lett.}\
  }\textbf {\bibinfo {volume} {85}},\ \bibinfo {pages} {2721} (\bibinfo {year}
  {2000})}\BibitemShut {NoStop}%
\bibitem [{\citenamefont {Chin}\ \emph {et~al.}(2004)\citenamefont {Chin},
  \citenamefont {Vuleti\'c}, \citenamefont {Kerman}, \citenamefont {Chu},
  \citenamefont {Tiesinga}, \citenamefont {Leo},\ and\ \citenamefont
  {Williams}}]{Chin:cs2-fesh:2004}%
  \BibitemOpen
  \bibfield  {author} {\bibinfo {author} {\bibfnamefont {C.}~\bibnamefont
  {Chin}}, \bibinfo {author} {\bibfnamefont {V.}~\bibnamefont {Vuleti\'c}},
  \bibinfo {author} {\bibfnamefont {A.~J.}\ \bibnamefont {Kerman}}, \bibinfo
  {author} {\bibfnamefont {S.}~\bibnamefont {Chu}}, \bibinfo {author}
  {\bibfnamefont {E.}~\bibnamefont {Tiesinga}}, \bibinfo {author}
  {\bibfnamefont {P.~J.}\ \bibnamefont {Leo}}, \ and\ \bibinfo {author}
  {\bibfnamefont {C.~J.}\ \bibnamefont {Williams}},\ }\href@noop {} {\bibfield
  {journal} {\bibinfo  {journal} {Phys. Rev. A}\ }\textbf {\bibinfo {volume}
  {70}},\ \bibinfo {pages} {032701} (\bibinfo {year} {2004})}\BibitemShut
  {NoStop}%
\bibitem [{\citenamefont {Berninger}\ \emph {et~al.}(2013)\citenamefont
  {Berninger}, \citenamefont {Zenesini}, \citenamefont {Huang}, \citenamefont
  {Harm}, \citenamefont {N\"agerl}, \citenamefont {Ferlaino}, \citenamefont
  {Grimm}, \citenamefont {Julienne},\ and\ \citenamefont
  {Hutson}}]{Berninger:Cs2:2013}%
  \BibitemOpen
  \bibfield  {author} {\bibinfo {author} {\bibfnamefont {M.}~\bibnamefont
  {Berninger}}, \bibinfo {author} {\bibfnamefont {A.}~\bibnamefont {Zenesini}},
  \bibinfo {author} {\bibfnamefont {B.}~\bibnamefont {Huang}}, \bibinfo
  {author} {\bibfnamefont {W.}~\bibnamefont {Harm}}, \bibinfo {author}
  {\bibfnamefont {H.-C.}\ \bibnamefont {N\"agerl}}, \bibinfo {author}
  {\bibfnamefont {F.}~\bibnamefont {Ferlaino}}, \bibinfo {author}
  {\bibfnamefont {R.}~\bibnamefont {Grimm}}, \bibinfo {author} {\bibfnamefont
  {P.~S.}\ \bibnamefont {Julienne}}, \ and\ \bibinfo {author} {\bibfnamefont
  {J.~M.}\ \bibnamefont {Hutson}},\ }\href@noop {} {\bibfield  {journal}
  {\bibinfo  {journal} {Phys. Rev. A}\ }\textbf {\bibinfo {volume} {87}},\
  \bibinfo {pages} {032517} (\bibinfo {year} {2013})}\BibitemShut {NoStop}%
\bibitem [{\citenamefont {Chevy}\ and\ \citenamefont
  {Salomon}(2016)}]{Chevy:2016}%
  \BibitemOpen
  \bibfield  {author} {\bibinfo {author} {\bibfnamefont {F.}~\bibnamefont
  {Chevy}}\ and\ \bibinfo {author} {\bibfnamefont {C.}~\bibnamefont
  {Salomon}},\ }\href@noop {} {\bibfield  {journal} {\bibinfo  {journal} {J.
  Phys. B}\ }\textbf {\bibinfo {volume} {49}},\ \bibinfo {pages} {192001}
  (\bibinfo {year} {2016})}\BibitemShut {NoStop}%
\bibitem [{\citenamefont {Kraemer}\ \emph {et~al.}(2006)\citenamefont
  {Kraemer}, \citenamefont {Mark}, \citenamefont {Waldburger}, \citenamefont
  {Danzl}, \citenamefont {Chin}, \citenamefont {Engeser}, \citenamefont
  {Lange}, \citenamefont {Pilch}, \citenamefont {Jaakkola}, \citenamefont
  {N\"{a}gerl},\ and\ \citenamefont {Grimm}}]{Kraemer:2006}%
  \BibitemOpen
  \bibfield  {author} {\bibinfo {author} {\bibfnamefont {T.}~\bibnamefont
  {Kraemer}}, \bibinfo {author} {\bibfnamefont {M.}~\bibnamefont {Mark}},
  \bibinfo {author} {\bibfnamefont {P.}~\bibnamefont {Waldburger}}, \bibinfo
  {author} {\bibfnamefont {J.~G.}\ \bibnamefont {Danzl}}, \bibinfo {author}
  {\bibfnamefont {C.}~\bibnamefont {Chin}}, \bibinfo {author} {\bibfnamefont
  {B.}~\bibnamefont {Engeser}}, \bibinfo {author} {\bibfnamefont {A.~D.}\
  \bibnamefont {Lange}}, \bibinfo {author} {\bibfnamefont {K.}~\bibnamefont
  {Pilch}}, \bibinfo {author} {\bibfnamefont {A.}~\bibnamefont {Jaakkola}},
  \bibinfo {author} {\bibfnamefont {H.~C.}\ \bibnamefont {N\"{a}gerl}}, \ and\
  \bibinfo {author} {\bibfnamefont {R.}~\bibnamefont {Grimm}},\ }\href@noop {}
  {\bibfield  {journal} {\bibinfo  {journal} {Nature}\ }\textbf {\bibinfo
  {volume} {440}},\ \bibinfo {pages} {315} (\bibinfo {year}
  {2006})}\BibitemShut {NoStop}%
\bibitem [{\citenamefont {Berninger}\ \emph {et~al.}(2011)\citenamefont
  {Berninger}, \citenamefont {Zenesini}, \citenamefont {Huang}, \citenamefont
  {Harm}, \citenamefont {N\"agerl}, \citenamefont {Ferlaino}, \citenamefont
  {Grimm}, \citenamefont {Julienne},\ and\ \citenamefont
  {Hutson}}]{Berninger:Efimov:2011}%
  \BibitemOpen
  \bibfield  {author} {\bibinfo {author} {\bibfnamefont {M.}~\bibnamefont
  {Berninger}}, \bibinfo {author} {\bibfnamefont {A.}~\bibnamefont {Zenesini}},
  \bibinfo {author} {\bibfnamefont {B.}~\bibnamefont {Huang}}, \bibinfo
  {author} {\bibfnamefont {W.}~\bibnamefont {Harm}}, \bibinfo {author}
  {\bibfnamefont {H.-C.}\ \bibnamefont {N\"agerl}}, \bibinfo {author}
  {\bibfnamefont {F.}~\bibnamefont {Ferlaino}}, \bibinfo {author}
  {\bibfnamefont {R.}~\bibnamefont {Grimm}}, \bibinfo {author} {\bibfnamefont
  {P.~S.}\ \bibnamefont {Julienne}}, \ and\ \bibinfo {author} {\bibfnamefont
  {J.~M.}\ \bibnamefont {Hutson}},\ }\href@noop {} {\bibfield  {journal}
  {\bibinfo  {journal} {Phys. Rev. Lett.}\ }\textbf {\bibinfo {volume} {107}},\
  \bibinfo {pages} {120401} (\bibinfo {year} {2011})}\BibitemShut {NoStop}%
\bibitem [{\citenamefont {Huang}\ \emph {et~al.}(2014)\citenamefont {Huang},
  \citenamefont {Sidorenkov}, \citenamefont {Grimm},\ and\ \citenamefont
  {Hutson}}]{Huang:2nd-Efimov:2014}%
  \BibitemOpen
  \bibfield  {author} {\bibinfo {author} {\bibfnamefont {B.}~\bibnamefont
  {Huang}}, \bibinfo {author} {\bibfnamefont {L.~A.}\ \bibnamefont
  {Sidorenkov}}, \bibinfo {author} {\bibfnamefont {R.}~\bibnamefont {Grimm}}, \
  and\ \bibinfo {author} {\bibfnamefont {J.~M.}\ \bibnamefont {Hutson}},\
  }\href@noop {} {\bibfield  {journal} {\bibinfo  {journal} {Phys. Rev. Lett.}\
  }\textbf {\bibinfo {volume} {112}},\ \bibinfo {pages} {190401} (\bibinfo
  {year} {2014})}\BibitemShut {NoStop}%
\bibitem [{\citenamefont {Fratini}\ and\ \citenamefont
  {Pieri}(2012)}]{Fratini:2012}%
  \BibitemOpen
  \bibfield  {author} {\bibinfo {author} {\bibfnamefont {E.}~\bibnamefont
  {Fratini}}\ and\ \bibinfo {author} {\bibfnamefont {P.}~\bibnamefont
  {Pieri}},\ }\href@noop {} {\bibfield  {journal} {\bibinfo  {journal}
  {Physical Review A}\ }\textbf {\bibinfo {volume} {85}},\ \bibinfo {pages}
  {063618} (\bibinfo {year} {2012})}\BibitemShut {NoStop}%
\bibitem [{\citenamefont {Tung}\ \emph {et~al.}(2014)\citenamefont {Tung},
  \citenamefont {Jimenez-Garcia}, \citenamefont {Johansen}, \citenamefont
  {Parker},\ and\ \citenamefont {Chin}}]{Tung:2014}%
  \BibitemOpen
  \bibfield  {author} {\bibinfo {author} {\bibfnamefont {S.-K.}\ \bibnamefont
  {Tung}}, \bibinfo {author} {\bibfnamefont {K.}~\bibnamefont
  {Jimenez-Garcia}}, \bibinfo {author} {\bibfnamefont {J.}~\bibnamefont
  {Johansen}}, \bibinfo {author} {\bibfnamefont {C.~V.}\ \bibnamefont
  {Parker}}, \ and\ \bibinfo {author} {\bibfnamefont {C.}~\bibnamefont
  {Chin}},\ }\href@noop {} {\bibfield  {journal} {\bibinfo  {journal} {Physical
  review letters}\ }\textbf {\bibinfo {volume} {113}},\ \bibinfo {pages}
  {240402} (\bibinfo {year} {2014})}\BibitemShut {NoStop}%
\bibitem [{\citenamefont {Gopalakrishnan}\ \emph {et~al.}(2015)\citenamefont
  {Gopalakrishnan}, \citenamefont {Parker},\ and\ \citenamefont
  {Demler}}]{Gopalakrishnan:2015}%
  \BibitemOpen
  \bibfield  {author} {\bibinfo {author} {\bibfnamefont {S.}~\bibnamefont
  {Gopalakrishnan}}, \bibinfo {author} {\bibfnamefont {C.~V.}\ \bibnamefont
  {Parker}}, \ and\ \bibinfo {author} {\bibfnamefont {E.}~\bibnamefont
  {Demler}},\ }\href@noop {} {\bibfield  {journal} {\bibinfo  {journal}
  {Physical review letters}\ }\textbf {\bibinfo {volume} {114}},\ \bibinfo
  {pages} {045301} (\bibinfo {year} {2015})}\BibitemShut {NoStop}%
\bibitem [{\citenamefont {Ardila}\ and\ \citenamefont
  {Giorgini}(2016)}]{Ardila:2016}%
  \BibitemOpen
  \bibfield  {author} {\bibinfo {author} {\bibfnamefont {L.~P.}\ \bibnamefont
  {Ardila}}\ and\ \bibinfo {author} {\bibfnamefont {S.}~\bibnamefont
  {Giorgini}},\ }\href@noop {} {\bibfield  {journal} {\bibinfo  {journal}
  {Physical Review A}\ }\textbf {\bibinfo {volume} {94}},\ \bibinfo {pages}
  {063640} (\bibinfo {year} {2016})}\BibitemShut {NoStop}%
\bibitem [{\citenamefont {DeSalvo}\ \emph {et~al.}(2019)\citenamefont
  {DeSalvo}, \citenamefont {Patel}, \citenamefont {Cai},\ and\ \citenamefont
  {Chin}}]{DeSalvo:2019}%
  \BibitemOpen
  \bibfield  {author} {\bibinfo {author} {\bibfnamefont {B.}~\bibnamefont
  {DeSalvo}}, \bibinfo {author} {\bibfnamefont {K.}~\bibnamefont {Patel}},
  \bibinfo {author} {\bibfnamefont {G.}~\bibnamefont {Cai}}, \ and\ \bibinfo
  {author} {\bibfnamefont {C.}~\bibnamefont {Chin}},\ }\href@noop {} {\bibfield
   {journal} {\bibinfo  {journal} {Nature}\ }\textbf {\bibinfo {volume}
  {568}},\ \bibinfo {pages} {61} (\bibinfo {year} {2019})}\BibitemShut
  {NoStop}%
\bibitem [{\citenamefont {Brue}\ and\ \citenamefont
  {Hutson}(2013)}]{Brue:AlkYb:2013}%
  \BibitemOpen
  \bibfield  {author} {\bibinfo {author} {\bibfnamefont {D.~A.}\ \bibnamefont
  {Brue}}\ and\ \bibinfo {author} {\bibfnamefont {J.~M.}\ \bibnamefont
  {Hutson}},\ }\href@noop {} {\bibfield  {journal} {\bibinfo  {journal} {Phys.
  Rev. A}\ }\textbf {\bibinfo {volume} {87}},\ \bibinfo {pages} {052709}
  (\bibinfo {year} {2013})}\BibitemShut {NoStop}%
\bibitem [{\citenamefont {Patel}\ \emph {et~al.}(2014)\citenamefont {Patel},
  \citenamefont {Blackley}, \citenamefont {Cornish},\ and\ \citenamefont
  {Hutson}}]{Patel:2014}%
  \BibitemOpen
  \bibfield  {author} {\bibinfo {author} {\bibfnamefont {H.~J.}\ \bibnamefont
  {Patel}}, \bibinfo {author} {\bibfnamefont {C.~L.}\ \bibnamefont {Blackley}},
  \bibinfo {author} {\bibfnamefont {S.~L.}\ \bibnamefont {Cornish}}, \ and\
  \bibinfo {author} {\bibfnamefont {J.~M.}\ \bibnamefont {Hutson}},\
  }\href@noop {} {\bibfield  {journal} {\bibinfo  {journal} {Phys. Rev. A}\
  }\textbf {\bibinfo {volume} {90}},\ \bibinfo {pages} {032716} (\bibinfo
  {year} {2014})}\BibitemShut {NoStop}%
\bibitem [{\citenamefont {Molony}\ \emph {et~al.}(2014)\citenamefont {Molony},
  \citenamefont {Gregory}, \citenamefont {Ji}, \citenamefont {Lu},
  \citenamefont {K\"oppinger}, \citenamefont {{Le Sueur}}, \citenamefont
  {Blackley}, \citenamefont {Hutson},\ and\ \citenamefont
  {Cornish}}]{Molony:RbCs:2014}%
  \BibitemOpen
  \bibfield  {author} {\bibinfo {author} {\bibfnamefont {P.~K.}\ \bibnamefont
  {Molony}}, \bibinfo {author} {\bibfnamefont {P.~D.}\ \bibnamefont {Gregory}},
  \bibinfo {author} {\bibfnamefont {Z.}~\bibnamefont {Ji}}, \bibinfo {author}
  {\bibfnamefont {B.}~\bibnamefont {Lu}}, \bibinfo {author} {\bibfnamefont
  {M.~P.}\ \bibnamefont {K\"oppinger}}, \bibinfo {author} {\bibfnamefont
  {C.~R.}\ \bibnamefont {{Le Sueur}}}, \bibinfo {author} {\bibfnamefont
  {C.~L.}\ \bibnamefont {Blackley}}, \bibinfo {author} {\bibfnamefont {J.~M.}\
  \bibnamefont {Hutson}}, \ and\ \bibinfo {author} {\bibfnamefont {S.~L.}\
  \bibnamefont {Cornish}},\ }\href@noop {} {\bibfield  {journal} {\bibinfo
  {journal} {Phys. Rev. Lett.}\ }\textbf {\bibinfo {volume} {113}},\ \bibinfo
  {pages} {255301} (\bibinfo {year} {2014})}\BibitemShut {NoStop}%
\bibitem [{\citenamefont {Takekoshi}\ \emph {et~al.}(2014)\citenamefont
  {Takekoshi}, \citenamefont {Reichs\"ollner}, \citenamefont {Schindewolf},
  \citenamefont {Hutson}, \citenamefont {{Le Sueur}}, \citenamefont {Dulieu},
  \citenamefont {Ferlaino}, \citenamefont {Grimm},\ and\ \citenamefont
  {N\"agerl}}]{Takekoshi:RbCs:2014}%
  \BibitemOpen
  \bibfield  {author} {\bibinfo {author} {\bibfnamefont {T.}~\bibnamefont
  {Takekoshi}}, \bibinfo {author} {\bibfnamefont {L.}~\bibnamefont
  {Reichs\"ollner}}, \bibinfo {author} {\bibfnamefont {A.}~\bibnamefont
  {Schindewolf}}, \bibinfo {author} {\bibfnamefont {J.~M.}\ \bibnamefont
  {Hutson}}, \bibinfo {author} {\bibfnamefont {C.~R.}\ \bibnamefont {{Le
  Sueur}}}, \bibinfo {author} {\bibfnamefont {O.}~\bibnamefont {Dulieu}},
  \bibinfo {author} {\bibfnamefont {F.}~\bibnamefont {Ferlaino}}, \bibinfo
  {author} {\bibfnamefont {R.}~\bibnamefont {Grimm}}, \ and\ \bibinfo {author}
  {\bibfnamefont {H.-C.}\ \bibnamefont {N\"agerl}},\ }\href@noop {} {\bibfield
  {journal} {\bibinfo  {journal} {Phys. Rev. Lett.}\ }\textbf {\bibinfo
  {volume} {113}},\ \bibinfo {pages} {205301} (\bibinfo {year}
  {2014})}\BibitemShut {NoStop}%
\bibitem [{\citenamefont {Esry}\ \emph {et~al.}(1999)\citenamefont {Esry},
  \citenamefont {Greene},\ and\ \citenamefont {Burke}}]{Esry:1999}%
  \BibitemOpen
  \bibfield  {author} {\bibinfo {author} {\bibfnamefont {B.~D.}\ \bibnamefont
  {Esry}}, \bibinfo {author} {\bibfnamefont {C.~H.}\ \bibnamefont {Greene}}, \
  and\ \bibinfo {author} {\bibfnamefont {J.~P.}\ \bibnamefont {Burke}},\
  }\href@noop {} {\bibfield  {journal} {\bibinfo  {journal} {Phys. Rev. Lett.}\
  }\textbf {\bibinfo {volume} {83}},\ \bibinfo {pages} {1751} (\bibinfo {year}
  {1999})}\BibitemShut {NoStop}%
\bibitem [{\citenamefont {Arndt}\ \emph {et~al.}(1997)\citenamefont {Arndt},
  \citenamefont {Ben~Dahan}, \citenamefont {Gu\'ery-Odelin}, \citenamefont
  {Reynolds},\ and\ \citenamefont {Dalibard}}]{Arndt:1997}%
  \BibitemOpen
  \bibfield  {author} {\bibinfo {author} {\bibfnamefont {M.}~\bibnamefont
  {Arndt}}, \bibinfo {author} {\bibfnamefont {M.}~\bibnamefont {Ben~Dahan}},
  \bibinfo {author} {\bibfnamefont {D.}~\bibnamefont {Gu\'ery-Odelin}},
  \bibinfo {author} {\bibfnamefont {M.~W.}\ \bibnamefont {Reynolds}}, \ and\
  \bibinfo {author} {\bibfnamefont {J.}~\bibnamefont {Dalibard}},\ }\href@noop
  {} {\bibfield  {journal} {\bibinfo  {journal} {Phys. Rev. Lett.}\ }\textbf
  {\bibinfo {volume} {79}},\ \bibinfo {pages} {625} (\bibinfo {year}
  {1997})}\BibitemShut {NoStop}%
\bibitem [{\citenamefont {Gu\'ery-Odelin}\ \emph {et~al.}(1998)\citenamefont
  {Gu\'ery-Odelin}, \citenamefont {S\"oding}, \citenamefont {Desbiolles},\ and\
  \citenamefont {Dalibard}}]{Guery-Odelin:1998}%
  \BibitemOpen
  \bibfield  {author} {\bibinfo {author} {\bibfnamefont {D.}~\bibnamefont
  {Gu\'ery-Odelin}}, \bibinfo {author} {\bibfnamefont {J.}~\bibnamefont
  {S\"oding}}, \bibinfo {author} {\bibfnamefont {P.}~\bibnamefont
  {Desbiolles}}, \ and\ \bibinfo {author} {\bibfnamefont {J.}~\bibnamefont
  {Dalibard}},\ }\href@noop {} {\bibfield  {journal} {\bibinfo  {journal}
  {Europhys. Lett.}\ }\textbf {\bibinfo {volume} {44}},\ \bibinfo {pages} {25}
  (\bibinfo {year} {1998})}\BibitemShut {NoStop}%
\bibitem [{\citenamefont {S\"oding}\ \emph {et~al.}(1998)\citenamefont
  {S\"oding}, \citenamefont {Gu\'ery-Odelin}, \citenamefont {Desbiolles},
  \citenamefont {Ferrari},\ and\ \citenamefont {Dalibard}}]{Soding:1998}%
  \BibitemOpen
  \bibfield  {author} {\bibinfo {author} {\bibfnamefont {J.}~\bibnamefont
  {S\"oding}}, \bibinfo {author} {\bibfnamefont {D.}~\bibnamefont
  {Gu\'ery-Odelin}}, \bibinfo {author} {\bibfnamefont {P.}~\bibnamefont
  {Desbiolles}}, \bibinfo {author} {\bibfnamefont {G.}~\bibnamefont {Ferrari}},
  \ and\ \bibinfo {author} {\bibfnamefont {J.}~\bibnamefont {Dalibard}},\
  }\href@noop {} {\bibfield  {journal} {\bibinfo  {journal} {Phys. Rev. Lett.}\
  }\textbf {\bibinfo {volume} {80}},\ \bibinfo {pages} {1869} (\bibinfo {year}
  {1998})}\BibitemShut {NoStop}%
\bibitem [{\citenamefont {Hopkins}\ \emph {et~al.}(2000)\citenamefont
  {Hopkins}, \citenamefont {Webster}, \citenamefont {Arlt}, \citenamefont
  {Bance}, \citenamefont {Cornish}, \citenamefont {Maragò},\ and\ \citenamefont
  {Foot}}]{Hopkins:2000}%
  \BibitemOpen
  \bibfield  {author} {\bibinfo {author} {\bibfnamefont {S.~A.}\ \bibnamefont
  {Hopkins}}, \bibinfo {author} {\bibfnamefont {S.}~\bibnamefont {Webster}},
  \bibinfo {author} {\bibfnamefont {J.}~\bibnamefont {Arlt}}, \bibinfo {author}
  {\bibfnamefont {P.}~\bibnamefont {Bance}}, \bibinfo {author} {\bibfnamefont
  {S.}~\bibnamefont {Cornish}}, \bibinfo {author} {\bibfnamefont
  {O.}~\bibnamefont {Maragò}}, \ and\ \bibinfo {author} {\bibfnamefont {C.~J.}\
  \bibnamefont {Foot}},\ }\href@noop {} {\bibfield  {journal} {\bibinfo
  {journal} {Phys. Rev. A}\ }\textbf {\bibinfo {volume} {61}},\ \bibinfo
  {pages} {032707} (\bibinfo {year} {2000})}\BibitemShut {NoStop}%
\bibitem [{\citenamefont {Thomas}\ \emph {et~al.}(2003)\citenamefont {Thomas},
  \citenamefont {Hopkins}, \citenamefont {Cornish},\ and\ \citenamefont
  {Foot}}]{Thomas:2003}%
  \BibitemOpen
  \bibfield  {author} {\bibinfo {author} {\bibfnamefont {A.~M.}\ \bibnamefont
  {Thomas}}, \bibinfo {author} {\bibfnamefont {S.}~\bibnamefont {Hopkins}},
  \bibinfo {author} {\bibfnamefont {S.~L.}\ \bibnamefont {Cornish}}, \ and\
  \bibinfo {author} {\bibfnamefont {C.~J.}\ \bibnamefont {Foot}},\ }\href@noop
  {} {\bibfield  {journal} {\bibinfo  {journal} {J. Opt. B}\ }\textbf {\bibinfo
  {volume} {5}},\ \bibinfo {pages} {S107} (\bibinfo {year} {2003})}\BibitemShut
  {NoStop}%
\bibitem [{\citenamefont {Weber}\ \emph {et~al.}(2003)\citenamefont {Weber},
  \citenamefont {Herbig}, \citenamefont {Mark}, \citenamefont {N\"{a}gerl},\
  and\ \citenamefont {Grimm}}]{Weber:CsBEC:2003}%
  \BibitemOpen
  \bibfield  {author} {\bibinfo {author} {\bibfnamefont {T.}~\bibnamefont
  {Weber}}, \bibinfo {author} {\bibfnamefont {J.}~\bibnamefont {Herbig}},
  \bibinfo {author} {\bibfnamefont {M.}~\bibnamefont {Mark}}, \bibinfo {author}
  {\bibfnamefont {H.~C.}\ \bibnamefont {N\"{a}gerl}}, \ and\ \bibinfo {author}
  {\bibfnamefont {R.}~\bibnamefont {Grimm}},\ }\href@noop {} {\bibfield
  {journal} {\bibinfo  {journal} {Science}\ }\textbf {\bibinfo {volume}
  {299}},\ \bibinfo {pages} {232} (\bibinfo {year} {2003})}\BibitemShut
  {NoStop}%
\bibitem [{\citenamefont {Yang}\ \emph {et~al.}(2019)\citenamefont {Yang},
  \citenamefont {Frye}, \citenamefont {Guttridge}, \citenamefont {\.Zuchowski},
  \citenamefont {Cornish},\ and\ \citenamefont {Hutson}}]{Yang:CsYb:2019}%
  \BibitemOpen
  \bibfield  {author} {\bibinfo {author} {\bibfnamefont {B.~C.}\ \bibnamefont
  {Yang}}, \bibinfo {author} {\bibfnamefont {M.~D.}\ \bibnamefont {Frye}},
  \bibinfo {author} {\bibfnamefont {A.}~\bibnamefont {Guttridge}}, \bibinfo
  {author} {\bibfnamefont {P.~S.}\ \bibnamefont {\.Zuchowski}}, \bibinfo
  {author} {\bibfnamefont {S.~L.}\ \bibnamefont {Cornish}}, \ and\ \bibinfo
  {author} {\bibfnamefont {J.~M.}\ \bibnamefont {Hutson}},\ }\href@noop {}
  {\bibfield  {journal} {\bibinfo  {journal} {arXiv:1905.xxxxx}\ } (\bibinfo
  {year} {2019})}\BibitemShut {NoStop}%
\bibitem [{\citenamefont {Cho}\ \emph {et~al.}(2013)\citenamefont {Cho},
  \citenamefont {McCarron}, \citenamefont {K\"oppinger}, \citenamefont
  {Jenkin}, \citenamefont {Butler}, \citenamefont {Julienne}, \citenamefont
  {Blackley}, \citenamefont {{Le Sueur}}, \citenamefont {Hutson},\ and\
  \citenamefont {Cornish}}]{Cho:RbCs:2013}%
  \BibitemOpen
  \bibfield  {author} {\bibinfo {author} {\bibfnamefont {H.-W.}\ \bibnamefont
  {Cho}}, \bibinfo {author} {\bibfnamefont {D.~J.}\ \bibnamefont {McCarron}},
  \bibinfo {author} {\bibfnamefont {M.~P.}\ \bibnamefont {K\"oppinger}},
  \bibinfo {author} {\bibfnamefont {D.~L.}\ \bibnamefont {Jenkin}}, \bibinfo
  {author} {\bibfnamefont {K.~L.}\ \bibnamefont {Butler}}, \bibinfo {author}
  {\bibfnamefont {P.~S.}\ \bibnamefont {Julienne}}, \bibinfo {author}
  {\bibfnamefont {C.~L.}\ \bibnamefont {Blackley}}, \bibinfo {author}
  {\bibfnamefont {C.~R.}\ \bibnamefont {{Le Sueur}}}, \bibinfo {author}
  {\bibfnamefont {J.~M.}\ \bibnamefont {Hutson}}, \ and\ \bibinfo {author}
  {\bibfnamefont {S.~L.}\ \bibnamefont {Cornish}},\ }\href@noop {} {\bibfield
  {journal} {\bibinfo  {journal} {Phys. Rev. A}\ }\textbf {\bibinfo {volume}
  {87}},\ \bibinfo {pages} {010703(R)} (\bibinfo {year} {2013})}\BibitemShut
  {NoStop}%
\bibitem [{\citenamefont {Hutson}\ and\ \citenamefont
  {Le~Sueur}(2019)}]{molscat:2019}%
  \BibitemOpen
  \bibfield  {author} {\bibinfo {author} {\bibfnamefont {J.~M.}\ \bibnamefont
  {Hutson}}\ and\ \bibinfo {author} {\bibfnamefont {C.~R.}\ \bibnamefont
  {Le~Sueur}},\ }\href {\doibase doi:10.1016/j.cpc.2019.02.014} {\bibfield
  {journal} {\bibinfo  {journal} {Comp. Phys. Comm.}\ }\textbf {\bibinfo
  {volume} {241}},\ \bibinfo {pages} {9} (\bibinfo {year} {2019})}\BibitemShut
  {NoStop}%
\bibitem [{\citenamefont {Manolopoulos}(1986)}]{Manolopoulos:1986}%
  \BibitemOpen
  \bibfield  {author} {\bibinfo {author} {\bibfnamefont {D.~E.}\ \bibnamefont
  {Manolopoulos}},\ }\href@noop {} {\bibfield  {journal} {\bibinfo  {journal}
  {J.~Chem. Phys.}\ }\textbf {\bibinfo {volume} {85}},\ \bibinfo {pages} {6425}
  (\bibinfo {year} {1986})}\BibitemShut {NoStop}%
\bibitem [{\citenamefont {Alexander}\ and\ \citenamefont
  {Manolopoulos}(1987)}]{Alexander:1987}%
  \BibitemOpen
  \bibfield  {author} {\bibinfo {author} {\bibfnamefont {M.~H.}\ \bibnamefont
  {Alexander}}\ and\ \bibinfo {author} {\bibfnamefont {D.~E.}\ \bibnamefont
  {Manolopoulos}},\ }\href@noop {} {\bibfield  {journal} {\bibinfo  {journal}
  {J.~Chem. Phys.}\ }\textbf {\bibinfo {volume} {86}},\ \bibinfo {pages} {2044}
  (\bibinfo {year} {1987})}\BibitemShut {NoStop}%
\bibitem [{\citenamefont {Hutson}(2007)}]{Hutson:res:2007}%
  \BibitemOpen
  \bibfield  {author} {\bibinfo {author} {\bibfnamefont {J.~M.}\ \bibnamefont
  {Hutson}},\ }\href@noop {} {\bibfield  {journal} {\bibinfo  {journal} {New J.
  Phys.}\ }\textbf {\bibinfo {volume} {9}},\ \bibinfo {pages} {152} (\bibinfo
  {year} {2007})}\BibitemShut {NoStop}%
\bibitem [{\citenamefont {Hutson}\ \emph {et~al.}(2009)\citenamefont {Hutson},
  \citenamefont {Beyene},\ and\ \citenamefont
  {Gonz\'{a}lez-Mart\'{\i}nez}}]{Hutson:HeO2:2009}%
  \BibitemOpen
  \bibfield  {author} {\bibinfo {author} {\bibfnamefont {J.~M.}\ \bibnamefont
  {Hutson}}, \bibinfo {author} {\bibfnamefont {M.}~\bibnamefont {Beyene}}, \
  and\ \bibinfo {author} {\bibfnamefont {M.~L.}\ \bibnamefont
  {Gonz\'{a}lez-Mart\'{\i}nez}},\ }\href@noop {} {\bibfield  {journal}
  {\bibinfo  {journal} {Phys. Rev. Lett.}\ }\textbf {\bibinfo {volume} {103}},\
  \bibinfo {pages} {163201} (\bibinfo {year} {2009})}\BibitemShut {NoStop}%
\end{thebibliography}%

\end{document}